\documentclass[sigconf,screen]{acmart}
\AtBeginDocument{%
  \providecommand\BibTeX{{%
    \normalfont B\kern-0.5em{\scshape i\kern-0.25em b}\kern-0.8em\TeX}}}

\acmConference[ACM e-Energy '23]{ACM E-Energy '23}{June 20--23,
  2023}{Orlando, FL}

\copyrightyear{2023}
\acmYear{2023}
\setcopyright{licensedusgovmixed}\acmConference[e-Energy '23]{The 14th ACM International Conference on Future Energy Systems}{June 20--23, 2023}{Orlando, FL, USA}
\acmBooktitle{The 14th ACM International Conference on Future Energy Systems (e-Energy '23), June 20--23, 2023, Orlando, FL, USA}
\acmPrice{15.00}
\acmDOI{10.1145/3575813.3595196}
\acmISBN{979-8-4007-0032-3/23/06}

\newcommand{\vect}[1]{\boldsymbol{\mathbf{#1}}}

\usepackage{caption}
\usepackage{subcaption}

\usepackage{accents}
\DeclareRobustCommand{\ubar}[1]{\underaccent{\bar}{#1}}

\begin{document}

\title{Managing Wildfire Risk and Promoting Equity through \\ Optimal Configuration of Networked Microgrids}

\author{Sofia Taylor}
\affiliation{%
  \institution{University of Wisconsin}
  \city{Madison}
  \state{Wisconsin}
  \country{USA}}

\author{Gabriela Setyawan}
\affiliation{%
  \institution{University of Wisconsin}
  \city{Madison}
  \state{Wisconsin}
  \country{USA}
}

\author{Bai Cui}
\affiliation{%
 \institution{National Renewable Energy Laboratory}
 \city{Golden}
 \state{Colorado}
 \country{USA}}

\author{Ahmed Zamzam}
\affiliation{%
  \institution{National Renewable Energy Laboratory}
 \city{Golden}
 \state{Colorado}
 \country{USA}}

\author{Line A. Roald}
\affiliation{%
  \institution{University of Wisconsin}
  \city{Madison}
  \state{Wisconsin}
  \country{USA}}

\begin{abstract}
    As climate change increases the risk of large-scale wildfires,
    wildfire ignitions from electric power lines are a growing concern. To mitigate the wildfire ignition risk, many electric utilities de-energize power lines to prevent electric faults and failures. These preemptive power shutoffs are effective in reducing ignitions, but they could result in wide-scale power outages.
    Advanced technology, such as networked microgrids, can help reduce the size of the resulting power outages;
    however, even microgrid technology might not be sufficient to supply power to everyone, thus forcing hard questions about how to prioritize the provision of power among customers. In this paper, we present an optimization problem that configures networked microgrids to manage wildfire risk while maximizing the power served to customers; however, rather than simply maximizing the amount of power served in kilowatts, our formulation also considers the ability of customers to cope with power outages, as measured by social vulnerability, and it discourages the disconnection of particularly vulnerable customer groups. %
    To test our model, we leverage a synthetic but realistic distribution feeder, along with publicly available social vulnerability indices and satellite-based wildfire risk map data, to quantify the parameters in our optimal decision-making model. Our case study results demonstrate the benefits of networked microgrids in limiting load shed and promoting equity during scenarios with high wildfire risk. 
\end{abstract}

\begin{CCSXML}
<ccs2012>
<concept>
<concept_id>10010405.10010481.10010484.10011817</concept_id>
<concept_desc>Applied computing~Multi-criterion optimization and decision-making</concept_desc>
<concept_significance>500</concept_significance>
</concept>
<concept>
<concept_id>10010583.10010662.10010668.10010671</concept_id>
<concept_desc>Hardware~Power networks</concept_desc>
<concept_significance>500</concept_significance>
</concept>
<concept>
<concept_id>10010583.10010662.10010668.10010672</concept_id>
<concept_desc>Hardware~Smart grid</concept_desc>
<concept_significance>500</concept_significance>
</concept>
</ccs2012>
\end{CCSXML}

\ccsdesc[500]{Applied computing~Multi-criterion optimization and decision-making}
\ccsdesc[500]{Hardware~Power networks}
\ccsdesc[500]{Hardware~Smart grid}

\keywords{networked microgrids, social vulnerability, wildfire risk, distribution systems, power shutoffs}

\begin{teaserfigure}
  \includegraphics[width=\textwidth]{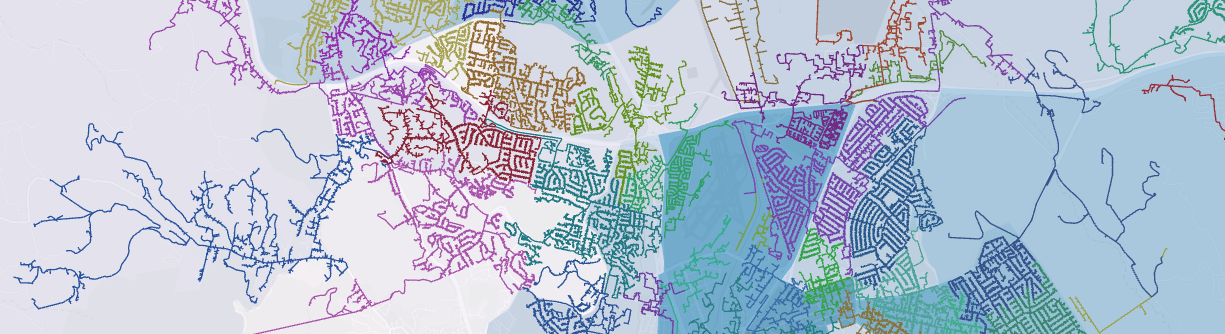}
  \caption{SMART-DS synthetic distribution system data \cite{SMART-DS} overlaid on top of the CDC Social Vulnerability Index map \cite{CDCSVI}.}
  \Description{SMART-DS synthetic distribution system data overlaid on top of the CDC Social Vulnerability Index map.}
  \label{fig:teaser}
\end{teaserfigure}

\received{10 February 2023}

\maketitle

\section{Introduction}
In the coming decades, our electric power systems must rapidly adapt to safely operate amid climate change-driven risks and extreme weather events
so that customers have access to a resilient supply of electricity. Further, we must also consider the social context surrounding electric power infrastructure to ensure equitable distribution of energy-related benefits and costs, especially in contingency scenarios.

A wide range of events can cause large-scale and sustained power outages. Networked microgrids \cite{Chen-NMGs} have recently been proposed as a technology that allows the grid to flexibly adapt and maintain operation during extreme scenarios. By enabling the dynamic formation of islanded grids supported by local generation, networked microgrids can provide power to local communities even in large-scale grid disruptions. 
In this paper, we consider how such microgrid technology can improve system resilience and promote equity in scenarios with high wildfire risk.

Wildfires create a complex environment for grid operations. Not only do wildfires threaten power infrastructure, but power equipment also has the potential to ignite fires through electric faults \cite{TAMU2014}.
To avoid such ignitions, electric grid operators implement preemptive power shutoffs, in which they de-energize grid components in high-risk areas and times, eliminating the potential for electric faults \cite{pgeMitigation}. While ``public safety power shutoffs'' successfully reduce ignitions, they can lead to widespread customer outages.
For example, a single shutoff event on October 9-12, 2019 avoided an estimated 114 ignitions, which might have impacted approximately 45,000 people and burned approximately 324,000 acres \cite{Technosylva2021october}; however, the shutoff caused 2.5 million people to lose power and over \$2 billion USD in economic losses \cite{Fuller2019pge, stanford2019pge}.
While power outages cause widespread economic and health effects \cite{anderson2020}, socially vulnerable people are disproportionately harmed by power outages \cite{Ham2022behavior, Sotolongo2020california}. Furthermore, research has demonstrated that disadvantaged communities are served by less advanced distribution grids, which limit access to, e.g., electric vehicle chargers and solar PV \cite{brockway2021inequitable}. 
To remedy these inequities, current government initiatives aim to direct government spending to disadvantaged communities \cite{Justice40}.

Considering this context, we examine the benefits of \emph{networked microgrids} \cite{Chen-NMGs, Wang-NMGs}, in equitably reducing wildfire ignition risks induced by power equipment.

\subsection{Related Work}
\subsubsection{Wildfire and Power System Interactions}
There are many mechanisms by which power lines can cause wildfire ignitions, including fallen power lines and poles, contact with vegetation or animals, malfunctioning equipment, or conductor slap \cite{russell2012distribution, TAMU2014, jazebi2020}.
Wildfire ignitions from power equipment are unfortunately not uncommon, and sometimes devastating. Prominent examples include the 2018 Camp Fire \cite{mohler2019cal} and the 2021 Dixie fire \cite{Moon2022california}, the deadliest and largest individual fires in California history, respectively. %
In general, fires started by power line faults often tend to be larger than fires from other ignition sources \cite{syphard2015location, victoria2009final}. 
This is likely because high wind speeds can lead to both a greater chance of failures of electric lines and a greater rate of fire spread.

To manage the risk of igniting wildfires by electrical components while minimizing the load shed due to power shutoffs, researchers have proposed several optimization-based decision-making models. The first optimization model to balance the trade-offs between minimizing wildfire risk and maximizing the load served is presented in \cite{rhodes2021balancing}. 
This optimal power shutoff problem was then extended in several works. The model in \cite{astudillo2022managing} formulates a multi-period optimal power shutoff problem, including representation of energy storage. %
In addition to selecting power shutoff locations, the rolling horizon optimization model in \cite{rhodes2022cooptimization} also determines how to restore power after shutoffs, given load and fire condition forecasts.

Aside from short-term grid operations, other efforts have considered long-term grid planning problems to mitigate wildfire risk. A framework to assess the wildfire risk of power lines and select overhead power lines to convert to underground cables is presented in \cite{taylor2021framework}. Also, the authors of \cite{kody2022optimizing} optimize line undergrounding, in conjunction with vegetation management and placement of distributed energy resources, to reduce ignition risk and power shutoffs.

\subsubsection{Networked Microgrids}
Conventionally, a microgrid is a static system that can be islanded from the rest of the electric grid to maintain power during blackouts \cite{Wang-NMGs}. Advances in remote switching and improved inverter control capabilities can enable microgrids to be dynamically formed within distribution networks to enhance the resiliency and make restoration more efficient. %
In general, for a microgrid to operate in isolation from the grid, it must include a grid-forming inverter to maintain the system's voltage and frequency. Other distributed generation sources are grid-following, and rely on this support to continue the supply of power.

Networked microgrids provide significant benefits in combatting the consequences of natural disasters. Grid resiliency against anticipated disasters has been shown to increase when networked microgrid approaches are used \cite{amirioun2017resilience, gholami2017proactive}. In addition, approaches that leverage networked microgrids have shown notable advantages in facilitating power restoration efforts \cite{arif-NMGs, Wang-NMGs}, and that the dynamic formation of community microgrids can enable more equitable restoration by leveraging customer-owned distributed energy resources  \cite{rhodes2021role}. 
In contrast, this paper focuses on the use of networked microgrids to manage the risks of igniting wildfires with a focus on equitable distribution of required power shutoffs.
Furthermore, while \cite{yang2022resilient} considers the use of microgrids to mitigate wildfire risk, it uses a much simpler formulation and does not model various microgrid capabilities or socioeconomic factors, as in this paper.

\subsubsection{Social Equity in Power Systems Decision-Making}

Although power outages are uniformly undesirable, certain social, environmental, and economic conditions can amplify impacts for socially vulnerable communities \cite{Ham2022behavior}.
\emph{Social vulnerability} describes a group's susceptibility to negative impacts from natural or human-caused hazards, including power outages, which is often a result of historical marginalization and underinvestment \cite{CDCSVI, Justice40}.
Vulnerability to power outages can arise from a range of intersecting factors.
Low-income customers might struggle to evacuate an area and find alternative accommodation, medically vulnerable or elderly customers might experience health conditions that are exacerbated without heating or air conditioning, and language and communication barriers might make it harder to elicit help. 
Recent studies of hurricane recoveries have also shown that certain indicators, such as minority status and residing in a rural location, are correlated with longer power outage durations and restoration times \cite{Mitsova_Esnard_Sapat_Lai_2018, Sotolongo_Kuhl_Baker_2021, Tormos2021}. Further, \cite{brockway2021inequitable} finds that Black-identifying and disadvantaged communities tend to live in communities where solar PV hosting capacity is low, thus limiting access to distributed generation.

In this work, we are interested in quantifying vulnerability to power outages so that grid operators can make more equitable decisions. 
One option is to use data sources that convey information about specific indicators.
One such tool is the U.S. Department of Health and Human Services' emPOWER map, which quantifies the number of individuals in a community that rely on electrically powered medical equipment \cite{HHSemPOWER}. 
Pacific Gas \& Electric also has a Medical Baseline program in which customers can indicate dependence on power for medical needs \cite{MedicalBaseline}.
Another option is to use data sources and mapping tools that aim to capture a broader definition of vulnerability by leveraging data in several different categories. %
Examples of such data sources include the Centers for Disease Control and Prevention (CDC) Social Vulnerability Index \cite{CDCSVI}, the U.S. Census' Community Resilience Estimates \cite{CensusCRE}, and the U.S. Council on Environmental Quality Climate and Economic Justice Screening Tool \cite{CEJST}. 
One limitation of such mapping tools is that a single ranking for a census tract does not capture the range of vulnerability, such as disabilities or reliance on electrically-powered medical equipment that might require special consideration during a power outage. They can, however, be useful to determine whether broader inequities exist, as demonstrated by the studies in \cite{Mitsova_Esnard_Sapat_Lai_2018, brockway2021inequitable}.  %

One distinguishing factor between our proposed work and prior work is the definition of fairness.
In general, equality emphasizes evenly distributed resources, whereas  
{equity} entails providing customized assistance based on individual needs \cite{Maeda_2019}. 
In the context of power shutoffs, \cite{kody2022sharing} utilizes an equality-based definition of fairness by proposing several cost functions for the equal distribution of power during shutoff events that
assume no knowledge of the social conditions of customers. 
In contrast, we define fairness in terms of equity, with electricity access for more vulnerable communities being prioritized in outage scenarios.

\subsection{Contributions}
The main contributions of this paper can be summarized as follows.

First, building on an existing model for the optimal operation of networked microgrids \cite{PMONM, Fobes23}, we propose extensions to incorporate the consideration of wildfire risk and to account for social vulnerability as part of the objective. We also discuss how the model can be adapted to consider different levels of grid controllability, including no microgrids, static microgrids with fixed borders, expanding microgrids (which can pick up additional load but not connect to other microgrids), and fully-flexible, networked microgrids. The resulting optimization problem, which we refer to as the \emph{optimal microgrid configuration problem}, allows us to assess the benefits of networked microgrids in terms of reducing wildfire risk and enabling equitable access to power across a range of scenarios.

Second, to inform our decision-making model, we develop a process to correlate real wildfire and social vulnerability data with the electric grid model. The method uses publicly available data sources for wildfire risk \cite{WFPI} and social vulnerability \cite{CDCSVI}, and combines them with realistic, but synthetic distribution feeder models, including one with 15,000 buses that is geographically located in California. Significant efforts went into processing the synthetic feeder model to reduce its size and to make it amenable to optimization. This test case and all associated data are made publicly available in \cite{OptMGConfigRepo}.

Third, we demonstrate our proposed optimal microgrid configuration problem in a realistic test case. We also perform a range of sensitivity studies to assess the impacts of different wildfire risk thresholds, objective function formulations with and without social vulnerability, and various microgrid technologies. The results demonstrate the benefits of networked microgrids in reducing wildfire risk, serving load, and promoting equity.

The remainder of the paper is organized as follows. Section \ref{sec:model} describes the modeling and optimization problem formulation. Section \ref{sec:dataprocess} describes the data sources and processing steps involved in developing the test case, and Section \ref{sec:results} describes the case study results. Finally, Section \ref{sec:conclusion} summarizes and concludes.

\section{Optimal Microgrid Configuration Problem}
\label{sec:model}

In this section, we present the optimal microgrid configuration problem, which is a mixed-integer linear programming problem. The system model is introduced first, followed by the introduction of the constraints and objective functions under different modeling assumptions. The overall problem formulation is presented last.

\subsection{System Modeling}
\label{sec:ops}

The optimization of switching decisions in distribution systems operations is implemented using the PowerModelsONM software package \cite{PMONM} and is referred to as the ``maximal load delivery'' (MLD) problem. The goal of this problem is to configure the topology to serve as much load (as measured by megawatts) as possible within the network constraints.
The main control variables are the switch statuses. 
Although PowerModelsONM is capable of determining  the optimal operations over multiple time steps, we focus on a single time period for simplicity.
In the remainder of this section, we review the key variables and constraints that are present in PowerModelsONM, as well as those that we have modified or added for this particular use case in the context of wildfire risk mitigation. To keep this model definition concise, we present generic formulations here and refer the reader to the PowerModelsONM documentation of the block-MLD problem \cite{PMONM} and the paper \cite{Fobes23} for exact formulations and more details on the power flow, operational, and topology constraints.

\subsubsection{System Model}

We consider a three-phase distribution system operating in steady state. The underlying topology of the system can be described by an undirected graph, $\mathcal{G} = (\mathcal{V}^+, \mathcal{E})$, where nodes and distribution lines are modeled as vertices, $\mathcal{V}^+$, and edges, $\mathcal{E} \subseteq \mathcal{V}^+ \times \mathcal{V}^+$, respectively. A subset of the distribution lines are switchable, meaning they can be either closed (1) or open (0). The switching status of line $(j,k) \in \mathcal{E}$ is denoted by $z^{sw}_{jk}$. The phases are denoted by $a, b$, and $c$ and are collected in $\Phi$, that is, $\Phi = \{ a, b, c \}$.

There is a single substation node in the system (node $0$), whose three-phase voltage magnitudes are assumed constant over the optimization horizon. We denote the set of nodes, other than the substation, by $\mathcal{V}$ so that $\mathcal{V}^+ = \mathcal{V} \cup \{0\}$. The nodes $i \in \mathcal{V}$ are modeled as constant power types whose real and reactive powers are regulated. Both loads and generators can be connected to a node, with power demand and generation denoted by $S^d_i = P^d_i + jQ^d_i$ and $S^g_i = P^g_i + jQ^g_i$, respectively. Some generators interface with the system through inverters, which have an operating mode, $z^{inv}$, that is either grid-forming (1) or grid-following (0).

\subsubsection{Load Blocks}
One important feature of the MLD problem in PowerModelsONM is the load block structure. Distribution grid loads typically cannot be controlled individually. Instead, load is served (or shed) through switching actions that energize (or de-energize) entire parts of the network, which we refer to as load blocks. Formally, load blocks are the connected components of the network when all switchable lines are open. We assume that all load blocks have an internally radial structure.  
Load block energization is denoted by a binary decision, variable, $z^{bl}_i$, in the optimal switching problem, where $z^{bl}_i=1$ indicates that load block $i$ is energized, and $z^{bl}_i=0$ indicates it is de-energized. 

\subsubsection{Power Flow Constraints}

The power flow is represented by a linear approximation of the unbalanced, three-phase power flow equations, known as LinDist3Flow, which was first proposed in \cite{arnold2016optimal}. The model describes a linear relationship between the squared nodal voltage magnitude difference between adjacent nodes and the nodal power injection. For line $(j,k) \in \mathcal{E}$, the linear power flow model is given by:
\begin{subequations}\label{eq:power_flow}
\begin{align}
    W_j &= W_k - \vect{M}_{jk}^P P_k - \vect{M}_{jk}^Q Q_k, \\
    \vect{M}_{jk}^P &= \begin{bmatrix} -2r^{aa}_{jk} & r^{ab}_{jk} - \sqrt{3}x^{ab}_{jk} & r^{ac}_{jk} + \sqrt{3}x^{ac}_{jk} \\ r^{ba}_{jk} + \sqrt{3}x^{ba}_{jk} & -2r^{bb}_{jk} & r^{bc}_{jk} - \sqrt{3}x^{bc}_{jk} \\
    r^{ca}_{jk} - \sqrt{3}x^{ca}_{jk} & r^{cb}_{jk} + \sqrt{3}x^{cb}_{jk} & -2r^{cc}_{jk}
    \end{bmatrix} \\
    \vect{M}_{jk}^Q &= \begin{bmatrix} -2x^{aa}_{jk} & x^{ab}_{jk} + \sqrt{3}r^{ab}_{jk} & x^{ac}_{jk} - \sqrt{3}r^{ac}_{jk} \\ x^{ba}_{jk} - \sqrt{3}r^{ba}_{jk} & -2x^{bb}_{jk} & x^{bc}_{jk} + \sqrt{3}r^{bc}_{jk} \\
    x^{ca}_{jk} + \sqrt{3}r^{ca}_{jk} & x^{cb}_{jk} - \sqrt{3}r^{cb}_{jk} & -2x^{cc}_{jk}
    \end{bmatrix} \\
    S_{ij} &= S_j + \sum_{(j,k) \in \mathcal{E}} S_{jk}
\end{align}
\end{subequations}
where $W_j$ is the squared three-phase voltage magnitude at node $j$, $r_{jk}^{\phi\varphi} + jx_{jk}^{\phi\varphi}$ is the mutual impedance of line $(j,k)$ between phases $\phi \in \Phi$ and $\varphi \in \Phi$ when $\phi \ne \varphi$, and the self impedance of phase $\phi$ when $\phi = \varphi$. Also, $S_j = S^d_j - S^g_j$ is the power consumption at node $j$ and $S_{ij}$ is the downstream line flow into node $j$.

\subsubsection{Operational Constraints}
There is also a set of operational constraints that enforces the component and system operating limits. Specifically, the following constraints are enforced:
\begin{subequations}\label{eq:operational_constraints}
\begin{align}
    & \ubar{V}_i \le V_i^\phi \le \bar{V}_i, && i \in \mathcal{V} \label{eq:op:v}\\
    & \ubar{S}_i^d \le  S_i^d \le \bar{S}_i^d, && i \in \mathcal{V} \label{eq:op:sd}\\
    & \ubar{S}_i^g \le S_i^g \le \bar{S}_i^g, && i \in \mathcal{V} \label{eq:op:sg}\\
    & \ubar{S}_{jk}^2\le P_{jk}^2 + Q_{jk}^2 \le \bar{S}_{jk}^2, && (j,k) \in \mathcal{E} \label{eq:op:sline}
\end{align}
\end{subequations}
where constraint \eqref{eq:op:v} bounds nodal voltage magnitudes, constraints \eqref{eq:op:sd} and \eqref{eq:op:sg} impose load and generator power limits, respectively, and constraint \eqref{eq:op:sline} enforces distribution line thermal limits.

\subsubsection{Topology Constraints}

The distribution system typically maintains a radial structure during normal operation for protection and ease of operation purposes. 
With grid-forming inverters, it is possible for part of the distribution system to island from the main grid to form microgrids. Each microgrid is constrained to contain exactly one grid-forming inverter and have a radial structure.

In PowerModelsONM, the multi-commodity flow formulation in conjunction with coloring constraints are used to ensure the connectivity and the existence and uniqueness of the grid-forming inverter in each microgrid \cite{PMONM}. The basic idea of the formulation is to assign a color to every grid-forming inverter, then assign the same color to every closed and energized switch in the same microgrid as the grid-forming inverter, and then ensure the existence of a path between the grid-forming inverter and the energized load block consisting of closed and energized switches of the same color. To ensure the system has no loops, an additional constraint is added that sets the number of energized lines to be one less than the difference between the number of energized nodes and the number of (energized) grid-forming inverters.

The full set of topology constraints is not covered here for brevity. Rather, we present a generic mixed-integer formulation as follows:
\begin{align}\label{eq:inv-1}
    \vect{A}\vect{z}^{inv} + \vect{B}\vect{z}^{sw} + \vect{C}\vect{z}^{bl} + \vect{D}\vect{y} \le \vect{b}
\end{align}
where $\vect{z}^{inv}$, $\vect{z}^{sw}$, and $\vect{z}^{bl}$ denote the operational mode of the inverters, status of switches, and energization of the load blocks, respectively. $\vect{A}, \vect{B}, \vect{C}, \vect{D}$ are constant matrices of appropriate dimensions, and $\vect{y} \in \mathbb{R}^{n_1} \times \{0,1\}^{n_2}$ is the vector of the auxiliary variables.

\subsection{Distribution Grid Controllability}

It is possible for distribution system operators to change the network topology through the operation of switches and inverters. 
In the optimal switching problem, the binary decision variable, $z^{sw}$, represent whether a switch is closed (1) or open (0),  and $z^{inv}$ represents whether an inverter is operating in a grid-forming (1) or grid-following (0) mode.
The values of $z^{sw}$ and $z^{inv}$ are optimized to satisfy the constraints and achieve an optimal objective function value. 
However, it is also possible to fix some of the $z^{sw}$ and $z^{inv}$ values to represent cases with less sophisticated microgrid control, that are more reflective of the current status of the grid. %
We define four levels of distribution grid controllability to assess how finer management of grid operations can achieve more benefits. 

\subsubsection{No Microgrids}
\label{sec:control1}
Our first case most closely reflects the current state of distribution grid operations, in which there are no microgrids in the network. This is equivalent to a situation where there are no distributed grid-forming inverters in the network and, thus, no microgrids can be formed. Any component must either have a path of connection to the substation or be de-energized.
To model this, we force all of the inverters to be grid following (i.e., the values of all of the $z^{inv}$ variables are equal to $0$), with the exception of the voltage source at the substation, which is grid forming. 
In this case, if there are component shutoffs due to high wildfire risk that sectionalize the grid, the radial structure of the distribution system would cause downstream load blocks to be shut off.

\subsubsection{Static Microgrids}
\label{sec:control2}
In this case, we allow static microgrids to form. This case reflects the current capabilities of conventional microgrids, in which microgrids are made up of a static set of components that can either all operate in grid-connected or islanded mode.
To do this, we manually set one inverter per load block to be grid-forming, while all others are set to be grid-following. 
We also set all switches to be open so that each load block with distributed generation and a grid-forming inverter can operate as a microgrid. 
In this setting, downstream load blocks may retain power even when upstream ones de-energize due to high wildfire risk.

\subsubsection{Expanding Microgrids}
\label{sec:control3}
This case models the situation in which static microgrids can form and also pick up additional load by expanding. However, two adjacent microgrids can not connect together (i.e. microgrids are not able to network). To model this, we again set one inverter per load block to be grid forming but we let the model choose the switch configurations.

\subsubsection{Networking Microgrids}
\label{sec:control4}
The final case, which reflects the given implementation of the optimal switching problem in PowerModelsONM, allows networking microgrids to form. All of $z^{sw}$ and $z^{inv}$ are free binary variables. 

\subsubsection{Summary of Distribution Grid Controllability}
\label{sec:control5}
The four cases are tabulated in Table \ref{tab:controllability} in terms of their respective admissible values of $z^{inv}$ and $z^{sw}$. For future reference, we denote the set of admissible values of $z^{inv}$ (resp. $z^{sw}$) in case $\ell$ ($\ell = 1, 2, 3, 4$) by $\mathcal{Z}^{inv}_\ell$ (resp. $\mathcal{Z}^{sw}_\ell$). For example, $\mathcal{Z}^{inv}_1 = \{0\}^{n^{inv}}$ and $\mathcal{Z}^{sw}_1 = \{0,1\}^{n^{sw}}$.

\begin{table}
    \captionsetup{justification=centering, labelsep=newline}
    \caption{\small Set of admissible values for every $z^{inv}$ and $z^{sw}$ under the four levels of distribution grid controllability}
    \label{tab:controllability}
    \centering
    \resizebox{\columnwidth}{!}{%
    \begin{tabular}{p{5cm}p{1.5cm}p{1.5cm}}
        \hline
         & $z^{inv}$ & $z^{sw}$ \\
        \hline
        No microgrids & $\{0\}$ & $\{0,1\}$ \\
        Static microgrids & $\{0\}$ or $\{1\}^\dagger$ & $\{0\}$ \\
        Expanding microgrids & $\{0\}$ or $\{1\}^\dagger$ & $\{0,1\}$ \\
        Networking microgrids & $\{0,1\}$ & $\{0,1\}$ \\
        \hline
        \multicolumn{3}{l}{\small{$^\dagger$ The value of $z^{inv}$ is either $0$ or $1$, depending on manual setting.}} \\
    \end{tabular}
    }
\end{table}

\subsection{Limiting Wildfire Ignition Risk}

Previous work modeled the mitigation of wildfire ignition risk as an objective function term, where load shed and wildfire risk reduction are competing objectives \cite{rhodes2021balancing}. However, in this work, we are interested in examining the definitions of load shed specifically, rather than balancing these objectives. Thus, we formulate wildfire risk mitigation as a constraint, where we specify an upper bound for the accepted level of system-wide wildfire ignition risk, as in

\begin{equation}
    \label{eq:riskcon}
    \frac{\mathcal{R}(\rho, z^{bl})} {R^{\text{total}}} \le \overline{R},
\end{equation}

\noindent where $R^{\text{total}}$ is the total possible wildfire ignition risk and $\overline{R}$ is a value between 0 and $1$ that indicates the maximum accepted fraction of wildfire risk. We also note that the constrained case can be viewed as adding the wildfire risk term to the objective with the corresponding optimal dual as the penalty parameter.

We define $\mathcal{R}$ as the sum of the wildfire ignition risk, $\rho_i$, for each load block $i$ that is energized, or 

\begin{equation}
    \label{eq:riskdef}
    \mathcal{R} = \sum_{i \in \mathcal{B}} \rho_{i} \cdot z^{bl}_{i}.
\end{equation}

In some cases, it may be important to consider the vulnerability of communities to wildfire ignitions in the above definition. 
However, in this work, we assume the community vulnerability to wildfires to be constant across all load blocks and enforce the constraint in Eq. \ref{eq:riskcon} with the risk definition in Eq. \ref{eq:riskdef}. Future work may study the hazards associated with potential wildfire ignitions in more detail.

\subsection{Equity-Aware Load Shed Modeling}
\label{sec:obj}

For our purposes, the objective function minimizes the load shed due to power shutoffs, which we define as a function of power demand, vulnerability, wildfire ignition risk, and a binary decision variable, $z^{bl}$, that indicates the energization of load blocks.

\begin{equation}
    \label{eq:objective}
    \min \quad  \mathcal{F}(P^d, v, z^{bl}).
\end{equation}

In this paper, we are interested in showing that the definition of this load shed function $\mathcal{F}$ matters. To assess this, we formulate three load shed cost functions: \emph{load only} cost, \emph{vulnerability only} cost, and \emph{vulnerability weighted} load cost.

\subsubsection{``Load Only'' Objective Function}
\label{sec:LOLS}

The load shed functon in Eq. \ref{eq:objective} is defined as

\begin{align}
    \label{eq:LOLS}
    \mathcal{F}_{\text{lo}}= \sum_{i \in \mathcal{B}} P^d_{i} \cdot (1 - z^{bl}_{i}).
\end{align}

\subsubsection{``Vulnerability Only'' Objective Function}
\label{sec:VOLS}
We also consider a formulation in which the demand, $P^D_{i}$, is not included and we instead consider only the customer vulnerability to power outages.

\begin{align}
    \label{eq:VOLS}
    \mathcal{F}_{\text{vo}} = \sum_{i \in \mathcal{B}} v^d_{i} \cdot (1 - z^{bl}_{i}).
\end{align}

With this formulation, the size of the demand does not influence the energization status of any particular load, i.e., loads with high consumption are not prioritized.
Factors like income might be correlated with electricity consumption in multiple ways. For example, high income customers might have larger homes and more electrically-powered machines and devices. However, these devices in high income households might also be more energy efficient. The \emph{vulnerability only} cost eliminates the possibility for these factors to skew load prioritization based on unequal energy consumption.

\subsubsection{``Vulnerability Weighted'' Objective Function}
\label{sec:VWLS}

Finally, we formulate a cost function that considers both the power demand and vulnerability of loads.

\begin{align}
    \label{eq:VWLS}
    \mathcal{F}_{\text{vl}} = \sum_{i \in \mathcal{B}} P^d_{i} \cdot v^d_{i} \cdot (1 - z^{bl}_{i}).
\end{align}

\subsection{Optimal Microgrid Configuration Problem}
After introducing the operational constraints and the equity-aware load shed models, we present the optimization formulation to be solved to obtain the optimal configuration of the distribution network. For a given cost function, $\mathcal{F}_{ob}$, where $ob \in \{\mathrm{lo},\mathrm{vo},\mathrm{vl}\}$ and level of distribution grid controllability, $\ell$, the optimal microgrid configuration problem can be described as follows:
\begin{subequations}
\begin{align}
    \textbf{OMCP}\text{:}\quad \min_{{\bf p}^d,\ {\bf V},\ {\bf z}^{bl},\ {\bf z}^{inv},\ {\bf z}^{sw}} &\quad  \mathcal{F}_{ob} (P^d, v, z^{bl}) \\
    \text{s.t.} &\quad \eqref{eq:power_flow}, \eqref{eq:operational_constraints}, %
    \eqref{eq:inv-1}\notag \\
    & \quad \vect{z}^{inv} \in \mathcal{Z}^{inv}_\ell, \vect{z}^{sw} \in \mathcal{Z}^{sw}_\ell \label{eq:OMCP:z} \\ 
    & \quad \frac{\mathcal{R}(\rho, z^{bl})} {R^{\text{total}}} \le \overline{R} \label{eq:limiting_risk}
\end{align}
\end{subequations}

\noindent The constraints \eqref{eq:power_flow} and \eqref{eq:operational_constraints} represent the power flow equations and the operational limits on system components. Equation \eqref{eq:inv-1} encapsulates the mixed-integer linear constraints that ensure the radiality of the distribution feeder and the existence and uniqueness of grid-forming inverter in each microgrid. Constraint \eqref{eq:OMCP:z} fixes certain inverter and switch statuses based on the level of distribution grid controllability. Notice that constraints \eqref{eq:power_flow}--\eqref{eq:operational_constraints} are enforced only for the energized elements. This can be modeled using disjunctive constraints, but the details are omitted for brevity. 
Finally, the constraint \eqref{eq:limiting_risk} limits the wildfire risk of the optimal solution to be below a specified threshold.

\section{Data Sources and Processing}
\label{sec:dataprocess}

\subsection{Electric Grid Data}
\label{sec:griddata}
This section describes the procedure to obtain data for the optimization model. The test case, including the electric grid data, wildfire risk data and social vulnerability data are publicly available in \cite{OptMGConfigRepo}.

\begin{figure}
     \centering
     \includegraphics[width=\linewidth]{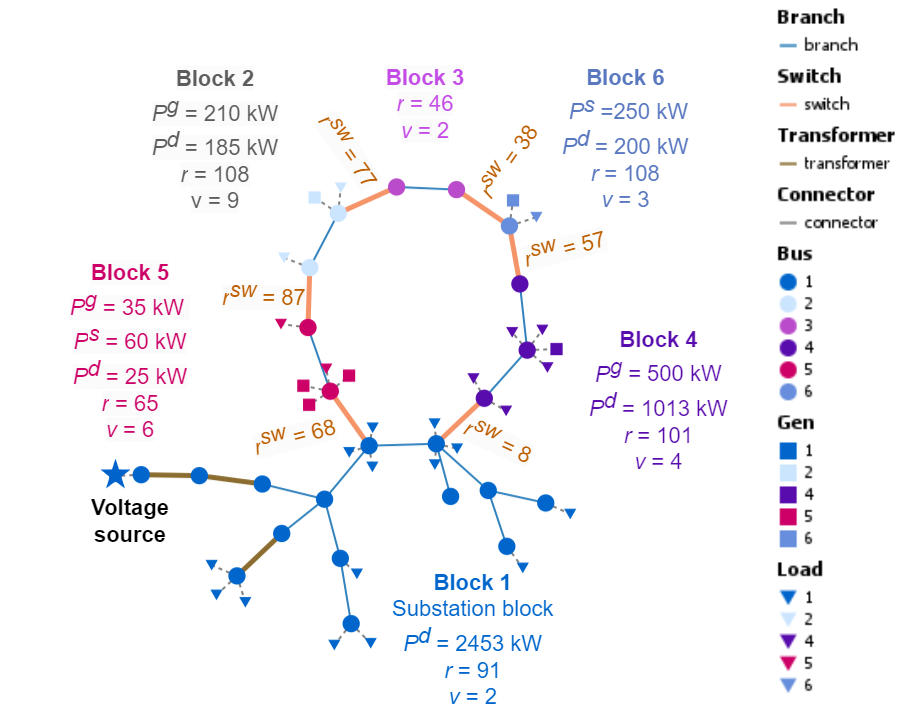}
    \caption{\small Modified IEEE 13-Bus feeder from PowerModelsONM \cite{PMONM}.}
    \label{fig:ieee13}
\end{figure}

\subsubsection{Data Source}
The problem is tested on two synthetic distribution grid models. The first is a modified version of the IEEE13 test case, which is available through PowerModelsONM package \cite{PMONM}. This test feeder is depicted in Fig. \ref{fig:ieee13}

The second is a modified version of a distribution system from the Synthetic Models for Advanced, Realistic Testing: Distribution Systems and Scenarios (SMART-DS) dataset \cite{SMART-DS}, which is published by the Grid Modernization group at the National Renewable Energy Laboratory. The grid models are created by utilizing the Reference Network Model \cite{domingo2011reference}, a tool developed by Universidad Pontificia Comillas, which designs a network topology to connect transmission substations, customer loads, and distributed energy resources via powerlines with paths that are geographically constrained by real street and building locations. The tool also adds components such as switches, transformers, and capacitors. All grid components are assigned geographic coordinates. 
The result is a set of realistic, but synthetic distribution grid models in locations including the San Francisco Bay Area.
For this test case, we choose five adjacent feeders, which all stem from a single substation (namely ``p17uhs13''), from the San Francisco Bay Area (``SFO'') network. 
These feeders are selected because they have sufficient geographic variation in both wildfire risk and social vulnerability indices, which are described in the following sections. 
We choose the version of this data with high solar and high battery penetration to give us flexibility in scaling generation and load.

\subsubsection{Feeder Reduction}
The SMART-DS feeder data has been modified so that it is compatible with the data parsing functions in the PowerModelsONM package \cite{PMONM}. Specifically, the functions are not able to parse the data's representation of the feeder configuration on the secondary side of the distribution transformers, i.e., home-level network details. 
Previous work in \cite{kroposki2020autonomous} reduced SMART-DS feeder data to only the primary circuit, aggregating the loads behind each distribution transformer and eliminating the secondary side. However, the data from \cite{kroposki2020autonomous} did not include information about PV systems, storage, or voltage regulators, since it utilized a previous version of the SMART-DS data. Therefore, we develop a similar method which we apply on the 2018 version of the SMART-DS data, which does contain those components. 

To reduce the considered feeders, we remove each distribution transformer one at a time, which allows us to identify the sub-graphs that contain the secondary side components. Within each secondary side sub-graph, we sum the real and reactive power values of all of the loads and solar PV units. These aggregated values are then assigned to the bus at the primary side of the distribution transformer. The energy storage units on the secondary side are not aggregated, but simply reflected to the buses at the primary side of the transformers. For all translated components, we are careful to maintain the same phase connections, so that the unbalanced nature of the system is still represented. We store the reduced data in a DSS file, which is publicly available in our data repository \cite{OptMGConfigRepo}.

\begin{figure}
     \centering
     \begin{subfigure}[b]{0.2\textwidth}
         \centering
         \includegraphics[width=\linewidth]{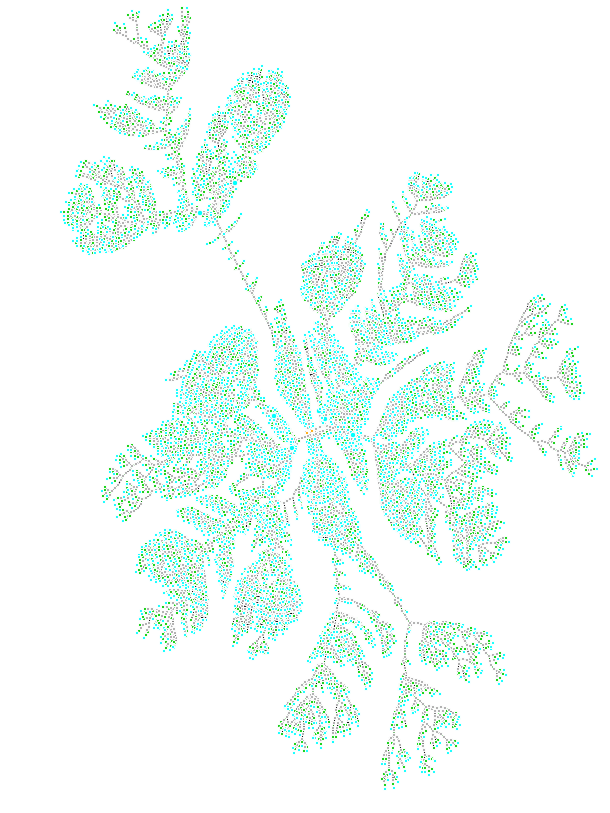}
         \caption{\small Original grid data.}
         \label{fig:orig_feeder}
     \end{subfigure}
     \hfill
     \begin{subfigure}[b]{0.26\textwidth}
         \centering
         \includegraphics[width=\linewidth]{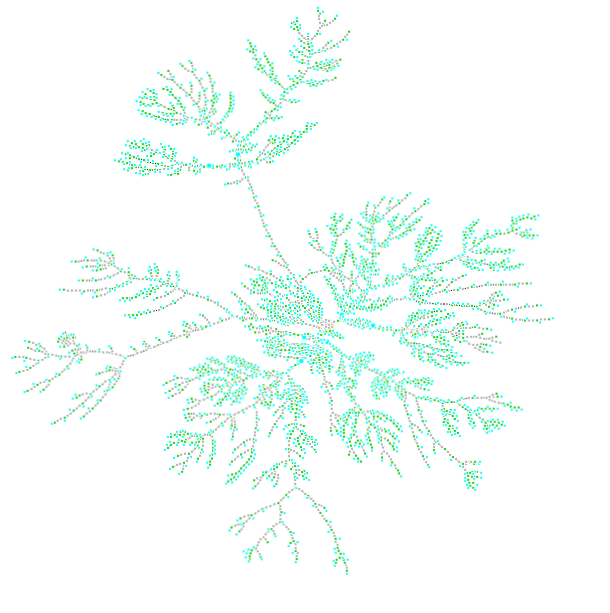}
         \caption{\small Reduced (modified) grid data.}
         \label{fig:smartds1}
     \end{subfigure}
        \caption{\small A portion of the SMART-DS grid data before and after modification. This corresponds to one substation, or five feeders, of synthetic distribution system data that is geo-located in the San Francisco Bay Area. }
        \label{fig:reduced_feeder}
\end{figure}

The original feeder data, shown in Fig. \ref{fig:orig_feeder}, has 15,299 buses, 12,639 lines, 14,639 loads, 4,611 solar PV units, 2,947 storage units, and 1920 transformers.
The reduced feeder data, in Fig. \ref{fig:reduced_feeder}, has 3,122 buses, 2,308 lines, 1,959 loads, 1,748 solar PV units, 2,947 storage units, and 74 transformers. 

\subsection{Wildfire Risk Data}

\subsubsection{Data Source}
To assign the wildfire risk values $\rho_{i}$ to each grid component $i$, we use the Wildland Fire Potential Index (WFPI) dataset. This dataset is represented in GIS maps that contain indices (between 0 and 150), which indicate the relative wildfire potential due to natural factors such as vegetation and weather conditions. 
Large fires and fire spread have historically occured at higher WFPI indices. The maps, which cover the contiguous U.S., are published once daily by the U.S. Geological Survey \cite{WFPI}. We select WFPI data from August 5, 2022, a day with relatively high wildfire risk. 

One limitation of this data is its spatial granularity of one square kilometer, which makes it somewhat challenging to discern differences in risk within a distribution-level system. Still, for the considered large-scale feeder, we observed reasonable variation in the wildfire risk indices across the different parts of the feeder. Further, there is only one set of wildfire data made available each day, which means that we cannot assess intra-day variations in risk. Despite these drawbacks, we note that the methods used to process the WFPI maps could also be applied to more granular wildfire risk data if such data were available.

\subsubsection{Processing and Aggregation Methods}
The WFPI data is available in a TIFF format, which we load into ArcGIS Pro, a GIS analysis application. 
First, we convert the wildfire risk map to a feature class. We overlay the wildfire risk feature class with the feeders from SMART-DS data. Finally, we extract values of wildfire risk indices for each SMART-DS line and export to a CSV file.

To be compatible with the optimization formulation in PowerModelsONM, which uses a load block structure and can not model secondary distribution circuits, we must aggregate the risk values from individual lines first to the primary circuit, and then to load blocks.
To do this, for each distribution transformer, we take the maximum risk value of all the secondary-side lines and assign it to the bus on the primary side of the transformer.
Finally, we assign the wildfire risk index of load block $i$, denoted by $\rho_{i}$, to be the maximum of the wildfire risk values of all components within the load block (which includes lines as well as buses at the primary side of distribution transformers).
The computed wildfire risk indices are then used in our optimization problem to represent the wildfire risk associated with energizing each load block.

\subsection{Social Vulnerability Data}
\label{sec:equitydata}

We need to quantify the social vulnerability parameters, $v_i$, which represent the vulnerability to power outages associated with the community within each load block $i$.
\subsubsection{Data Source}
We choose the CDC SVI to quantify the vulnerability to power outages, $v_i$, due to its granularity and ability to capture different aspects of vulnerability. The SVI represents the relative social vulnerability of each census tract based on underlying demographic indicators. It ranks each tract into 16 social factors and groups into 4 related themes. The themes are socioeconomic status, household composition and disability, minority status and language, and housing type and transportation. In addition, the CDC SVI also provides an overall ranking by summing up the ranking for each theme, ordering them and calculating overall percentile rankings.  

\subsubsection{Processing and Aggregation Methods}
The SVI data is imported to ArcGIS Pro in a SHP file. We use the most recent SVI data published by CDC.
We overlay the selected SMART-DS feeder with the SVI dataset on a map to each load in the feeder to its associated SVI values. 
The resulting assigned SVI values are exported as a CSV file, similar to the approach we used for the WFPI data. %

To be compatible with the optimization formulation in PowerModelsONM, we must aggregate the vulnerability values from individual loads first to the primary circuit, and then to load blocks. We sum the SVI values for all load on the secondary circuits and assign these values to the corresponding bus on the primary side.
Then, we sum the SVI values from individual buses to load blocks
to quantify the vulnerability of load block $i$, denoted by $v_i$. These values are used as weights in terms of optimization cost function to promote more equitable solutions.

\section{Test Case Results}
\label{sec:results}

We assess how the optimal configuration of networked microgrids can help distribution utilities manage wildfire risk and promote equitable access to electricity during disruptive events. The case study contains two parts, outlined below.

First, we implement the optimal microgrid configuration problem for the SMART-DS test case to demonstrate the use of the method on a realistic, large-scale system with real wildfire risk and community vulnerability data. 
Second, we perform a sensitivity analysis of the model where we assess the impact of different wildfire risk thresholds and investigate how the use of different objective functions (reflecting different ways of modeling impact of power outages on the community) and different microgrid controllability levels (ranging from no microgrids to fully dynamic, networked microgrid configurations) impact the optimal solution. Due to the significant computational overhead associated with solving the problem for the SMART-DS test case, we use a smaller model based on the IEEE13 test case for these sensitivity analyses. 

\subsection{SMART-DS Test Case Results}
In this example, we use the SMART-DS test case described in the section above to evaluate the performance of our model on a large, realistic case with real wildfire and social vulnerability data. We assume that the grid can be operated in a networked microgrid configuration that minimizes the \emph{vulnerability weighted} load shed cost, and choose a wildfire risk upper bound of 50\%. 
Since we originally selected the SMART-DS data that has high penetration of distributed solar and battery resources, we found that there is a lot more generation than load in the system. To create a realistic case where not all renewable generation is able to produce 100\% of their rated power and the disruption happens at a time where not all storage facilities are full, we downscale the available power of each distributed generation source and storage unit to 20\% of their original values. 
After this change, the total distributed generation in the network is 6.631 MW, the total storage capacity is 5.310 MW, and the total load is 17.012 MW. 
This particular instance of the model took approximately 70 minutes to solve on a personal computer using the commercial solver, Gurobi. 

In our results, we observe that 363 blocks are energized (out of 735), while only 12 out of 734 switches are closed. Most of the de-energized load blocks do not contain load, so despite the large number of de-energized blocks our solution still serves 16,952 kW or 99.6\% of the total. In terms of \emph{vulnerability-weighted} load value, which is the objective function we used, we serve 18,822.3 out of 18,822.9 or 99.996\% of the total. The \emph{vulnerability only} objective for this solution is 47.4 (or 92.7\% of the total).
The average vulnerability index of the loads served is 0.13, while the average vulnerability index of the loads that are not served is 0.01. 
The total wildfire risk is reduced to 49.9\% of the original.

We conclude that the system -- due to the large number of distributed energy resources -- is able to serve almost all the load even when most switches are open. However, the few switches that are closed indicate that the presence of networked microgrid capabilities (where some load blocks are able to connect to each other) is beneficial to the solution. Overall, the ability to operate the grid in networked microgrid configuration enables significant reductions in wildfire risk without sacrificing load served, both in absolute terms and when considering social vulnerability. 

\subsection{Sensitivity Analysis}

Next, we perform a more in-depth comparison of the different parameter choices and problem formulations.

\subsubsection{IEEE13 Test Case}
The subsequent sensitivity analyses require running many test cases. To keep computational effort manageable, we use the smaller IEEE13 test system for these comparisons. The original IEEE13 system is available at \cite{IEEE13orig}, but we use the modified version in the PowerModelsONM Julia package \cite{PMONM}. 
Further modifications to the test system for specific cases are described in the following subsections.
Since the IEEE13 case has no geographic information, the wildfire risk and vulnerability values are randomly assigned. The risk values for load blocks and switches are randomly generated values between 0 and 150, and 0 and 100, respectively. Vulnerability indices for load blocks are randomly generated values between 0 and 10. These values are shown in Fig. \ref{fig:ieee13} and the full datasets are available in our code repository \cite{OptMGConfigRepo}.

\subsubsection{Illustrative Example}
In this example, we assess the effect of controlling the level of acceptable wildfire risk on the resulting solution. %
We assume that the grid is able to support networked microgrid configurations that minimize the \emph{vulnerability weighted} load shed cost for a given wildfire risk upper bound equal to 0.5, i.e., we are willing to accept up to 50\% of the wildfire ignition risk.

\begin{figure}
    \centering
    \includegraphics[width=\linewidth]{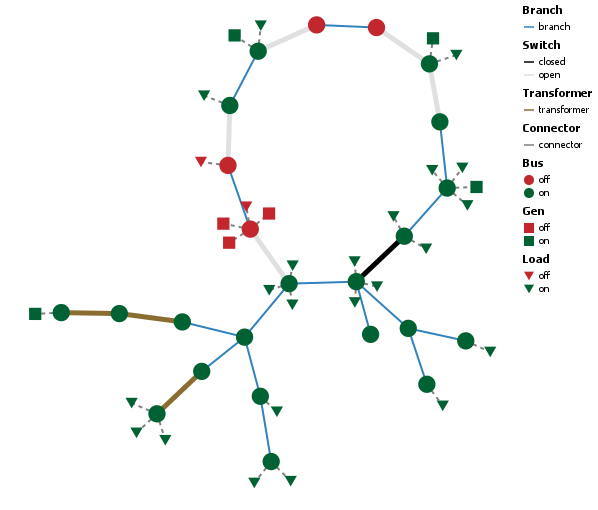}
    \caption{\small Plot of modified IEEE13 system depicting energization of components in the illustrative example.}
    \label{fig:energization}
\end{figure}

For this case, we observe the following results. 
Out of 6 load blocks, four blocks (1, 2, 4, and 6) are energized and two (3 and 5) are de-energized. Note that Block 1 contains the substation and is always energized.
Out of 6 switches, only switch 2 is closed, while the rest are open. 
The block energization and switch statuses are shown in Fig. \ref{fig:energization}. 
As a result of this configuration, the resulting wildfire risk is 416 out of a the system-wide potential of 854, or 48.7\%, which is achieved through shutoffs. 
The \emph{vulnerability-weighted} load served is 11,223 kW out of 11,373 kW, or 98.7\%. 
The load served (without vulnerability weighting) is 3,851 out of 3,876 kW, or 99.4\%. 
The \emph{vulnerability only} load served is 18 out of 26, or 69.2\%.
The average vulnerability index of the loads served is 4.5, while the average vulnerability index of the loads that are not served is 4.0. 

In this illustrative example, it is interesting that nearly all of the load can be served while reducing the wildfire risk by half. We also note that the model tends to open many switches. The switches have wildfire risk values associated with them, so de-energizing the switches can achieve large reductions in wildfire risk without necessarily shedding load, provided that there is sufficient distributed generation. However, the switch between load blocks 1 and 4 remains closed in this example, such that the power injection from the substation in block 1 can pick up the large load in block 4.

\subsubsection{Sensitivity to Wildfire Risk Threshold}
\label{sec:sens_to_threshold}
By varying the value of the accepted wildfire risk threshold, we can find different solutions with different levels of load shed and wildfire risk. 
To generate these results, we run the optimal microgrid configuration problem with \emph{vulnerability weighted} load shed and networked microgrid capability while varying the accepted risk threshold between 0 and 1 in steps of 0.001.

\begin{figure}
    \centering
    \includegraphics[width=\linewidth]{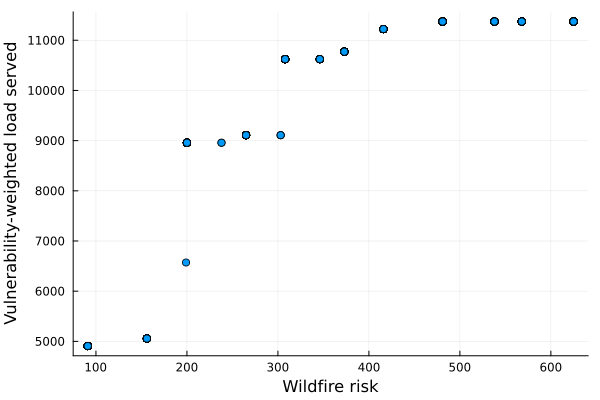}
    \caption{\small Minimum \emph{vulnerability weighted} load values obtained by varying the total accepted wildfire risk for the IEEE13 test case.}
    \label{fig:solns}
\end{figure}

These solutions are depicted in Fig. \ref{fig:solns}. As expected, the amount of \emph{vulnerability weighted} load served increases as we accept larger wildfire risk values (corresponding to a relaxation in the constraints of the optimization problem). Interestingly, despite the large number of risk thresholds tested here, there are only 15 unique solutions for this small test case. We also note that there are ``jumps'' in \emph{vulnerability weighted} load served between different solutions.

The small number of solutions and the sudden increases in wildfire risk can be attributed to the small number load blocks in the IEEE13 test case and the large variance in their sizes. We note that running a sensitivity analysis is quite useful in determining the best solutions for any given situation. For example, if a distribution grid operator is willing to accept a system-wide wildfire risk of approximately 200, they can increase in \emph{vulnerability weighted} load served for a very small increase in accepted wildfire risk.

\subsubsection{Impact of Equity Considerations on Load Serving Prioritization}
Next, we demonstrate how different formulations of the load shed objective, introduced in Section \ref{sec:obj}, impact how load blocks are prioritized in shutoff decisions.
To determine this order, we run the optimization problem many times, change the upper limit of the accepted wildfire risk from 0 to 1 by steps of 0.001, as done in Section \ref{sec:sens_to_threshold},  recording which load blocks are on at each step. The highest-priority block (labeled 1\textsuperscript{st}) is the one that is on for most of these steps, and so on.
Note that the load block shutoff ordering is not always sequential, i.e., a load block could be de-energized at a particular risk threshold, and then turned back on at a lower threshold.
Also note that Block 1 contains the substation and is always energized, so it is not considered in the priority ordering.

Table \ref{tab:priority} shows the prioritization of serving particular load blocks across the different wildfire risk thresholds and for different objective function. 
When minimizing the \emph{load only} or \emph{vulnerability only} objective functions, the priorities generally align with the magnitude of the load blocks' kilowatt load amount and vulnerability index, respectively. 
There are exceptions to this trend that arise due to the model's constraints (power flow, wildfire risk limits, etc.). 
The placement changes (in parentheses) show that there is large difference in the order between the \emph{load only} and \emph{vulnerability only} costs; however, the changes between both of  these and the \emph{vulnerability weighted} load cost are smaller, indicating that the  \emph{vulnerability weighted} objective identifies an intermediate solution that considers both the kilowatt load and the vulnerability index. 

This comparison indicates that the  \emph{vulnerability-weighted} load cost is effective in selecting microgrid configurations that prioritize serving large loads and vulnerable loads. We also observe that the common approach of using only the kilowatt load to determine shutoff priorities might leave particularly vulnerable communities at risk. For example, the  \emph{load only} cost places block 2, the most vulnerable load block, fourth of six load blocks.

\begin{table*}[ht]
    \captionsetup{justification=centering, labelsep=newline}
    \caption{\small Prioritization of serving load blocks in the modified IEEE13 system for the three load-serving cost functions.}
    \label{tab:priority}
    \centering
    \resizebox{\textwidth}{!}{%
        \begin{tabular}{p{1.2cm}p{1.2cm}p{1.2cm}p{1.5cm}p{3cm}p{1.65cm}p{1.75cm}p{2.95cm}}
            \hline
            & & & & & \multicolumn{3}{c}{Shutoff Priority (Change Relative to Load Only)} \\
            \cline{6-8}
            Load Block & Wildfire Risk & Load (kW)  & Vulnerability Index &   Vulnerability Weighted Load Value &   Load Only Cost &   Vulnerability Only Cost  &  Vulnerability Weighted Load Cost\\
            \hline
            1 & 91 & 2453    & 2 & 4.906 & N/A  & N/A  & N/A  \\% & N/A  \\
            2 & 108 & 185     & 9 & 1.665 & 4\textsuperscript{th}  & 1\textsuperscript{st} {\color{green} (+3)} & 2\textsuperscript{nd} ({\color{green} +2} %
            ) \\% & \\
            3 & 46 & 0       & 2 & 0     & 5\textsuperscript{th}  & 4\textsuperscript{th} {\color{green} (+1)} & 5\textsuperscript{th} ({\color{blue}+0} %
            ) \\% & \\
            4 & 101 & 1013    & 4 & 4.052 & 1\textsuperscript{st}  & 3\textsuperscript{rd} {\color{red} (-2)} & 1\textsuperscript{st} ({\color{blue}+0} %
            ) \\% & \\
            5 & 65 & 25      & 6 & 0.15  & 3\textsuperscript{rd}  & 2\textsuperscript{nd} {\color{green} (+1)} & 3\textsuperscript{rd} ({\color{blue}+0}%
            ) \\% & \\
            6 & 108 & 200     & 3 & 0.6   & 2\textsuperscript{nd}  & 5\textsuperscript{th} {\color{red} (-3)} & 4\textsuperscript{th} ({\color{red}-2}
            ) \\% & \\
            \hline
        \end{tabular}
    }
\end{table*}

\subsubsection{Impact of Networked Microgrids in Local Wildfire Risk Scenario}

We next consider the scenario where there is a high wildfire risk locally in the community, but the substation is still connected to the main grid and able to serve electricity to the feeder. We seek to assess how the presence of microgrids with networking capabilities (as described in Section \ref{sec:control4}) compares with a normal distribution feeder, where a connection to the substation is required to maintain energization status (as described in Section \ref{sec:control1}). 
We again run the optimization problem for the modified IEEE13 system, where the objective function is defined using the   \emph{vulnerability weighted} load shed cost \eqref{eq:VWLS}. We set the acceptable wildfire risk level to 50\%.
The results for this comparison are shown in Table \ref{tab:ieee13_control1}.

In this table, we observe that the case with networked microgrids keeps more load blocks on and fewer switches closed compared with the standard distribution grid case with no grid-forming inverters. Both solutions maintain a wildfire risk level below the acceptable level of 50\%, but the networked microgrids case is able to serve 98\% of the load, which is 6\% higher than the standard distribution grid case. In addition to these results, we also list the average SVI for both the loads that are served and the loads that are shed. We see that the vulnerability of the loads served by the networked microgrids case is higher than for the standard distribution grid case (i.e., more vulnerable populations are provided access to electricity), and the vulnerability of the loads that are shed is lower (i.e., less vulnerable populations experience outages).
Overall, these results demonstrate that the networked microgrids paradigm enables reducing the size of power outages due to preemptive power shutoffs, and better prioritizes service to vulnerable populations.

\begin{table}
    \captionsetup{justification=centering, labelsep=newline}
    \caption{\small Controllability comparison of no microgrids versus  networked microgrids for a wildfire risk threshold of 50\%. }
    \label{tab:ieee13_control1}
    \centering
    \resizebox{0.9\columnwidth}{!}{%
    \begin{tabular}{p{4.5cm}p{2cm}p{1.5cm}}
        \hline
         & No \linebreak {Microgrids} & Networking Microgrids \\
        \hline
        Load blocks on (out of 6) & 3 & 4 \\
        Switches closed (out of 6) & 3 & 1 \\
        Wildfire risk accepted & 47.7\% & 46.8\%  \\
          Vulnerability-weighted load served & 92.2\% & 98.0\% \\
        Vulnerability of load served & 4 & 5.25 \\
        Vulnerability of load not served & 4.67 & 2.5 \\
        \hline
    \end{tabular}
    }
\end{table}

\subsubsection{Impact of Networked Microgrids in Widespread Wildfire Risk Scenario}
\label{sec:widespread}

Finally, we consider a scenario where there is high wildfire risk throughout the region and the substation is de-energized as part of a larger public safety power shutoff. 
For this case, we compare three microgrid settings, including static microgrids, expanding microgrids, and networking microgrids, described in Sections \ref{sec:control2},  \ref{sec:control3}, and \ref{sec:control4}, respectively.
We model the substation disconnection by enforcing the switch connecting the voltage source to the rest of the IEEE13 system to be always open. 
As the IEEE13 case is small and does not lead to many unique solutions, we also adjust the load values in three of the load blocks to better illustrate the performance of different microgrid capabilities, as in Table \ref{tab:ieee13_changes}.
\begin{table}
    \captionsetup{justification=centering, labelsep=newline}
    \caption{\small Changes to demand values for the IEEE13 load blocks in Sec. \ref{sec:widespread}}
    \label{tab:ieee13_changes}
    \centering
    \resizebox{0.9\columnwidth}{!}{%
    \begin{tabular}{p{3cm}p{0.5cm}p{0.5cm}p{0.5cm}p{0.5cm}p{0.5cm}p{0.5cm}}
        \hline
        Load block & 1 & 2 & 3 & 4 & 5 & 6 \\
        \hline
        Original demand (kW) &  2453 & 185 & 0 & 1013 & 25 & 200\\
        Modified demand (kW) & 2453 & 185 & 10 & 405.2 & 25 & 260 \\
        \hline
    \end{tabular}
    }
\end{table}
We run the optimization problem where the objective function minimizes the   \emph{vulnerability weighted} load shed cost and the wildfire risk threshold is 90\%. The results for this comparison are in Table \ref{tab:ieee13_control2}.

As the level of controllability increases from static to expanding and networking microgrids, we observe that the number of energized load blocks and closed switches increases and the amount of   \emph{vulnerability weighted} load served increases. The increase in load served is most pronounced when we consider networked microgrids, as several microgrids with excess generation combine to serve additional load. All the cases maintain a wildfire risk well below the threshold of 90\%, since their ability to energize the load blocks is limited by their available generation capacity and their ability to connect to neighboring load blocks with excess generation. 

When analyzing this case in more detail, we see that in the static microgrid case, load blocks 2, 4, and 5 are energized, because they have sufficient generation to meet their load. In the expanding microgrid case, the switch between blocks 2 and 3 closes, because block 2 has enough excess generation to pick up the 10 kW load at block 3. As a result, blocks 2, 3, 4, and 5 are energized.
In the networking microgrids case, the blocks 2, 3, and 5 are connected to one bigger microgrid, which is also able to energize to block 6.

When comparing the vulnerability of the load served and not served for the three microgrid cases, we see that vulnerability levels decrease with increasing controllability. This is because the cases with more controllability enable serving more load.

\begin{table}
    \captionsetup{justification=centering, labelsep=newline}
    \caption{\small Controllability comparison of static, expanding, and networked microgrids for a wildfire risk threshold of 90\%}
    \label{tab:ieee13_control2}
    \centering
    \resizebox{\columnwidth}{!}{%
    \begin{tabular}{p{4.45cm}p{1.4cm}p{1.45cm}p{1.5cm}}
        \hline
         &    Static \linebreak {microgrids} & Expanding microgrids & Networking microgrids \\
        \hline
        Load blocks on (out of 7)  & 3 & 4 & 5  \\
        Switches closed (out of 6)  & 0 & 1 & 3 \\
        Wildfire risk accepted & 32.08\% & 46.49\% & 73.77\% \\
        Vulnerability-weighted load served & 34.69\% & 34.92\%  & 43.85\% \\
        Vulnerability of load served & 6.33 & 5.25 & 4.8  \\
        Vulnerability of load not served  & 1.75 & 1.67 & 1.0 \\
        \hline
    \end{tabular}
    }
\end{table}

\section{Conclusion}
\label{sec:conclusion}

Adapting to climate change and ensuring equitable outcomes for customers are two of the main challenges facing the power systems community today. Conventionally, electric grid decision-making models do not consider these factors. To address this gap, this paper incorporates environmental and demographic data into optimization models to capture the real setting of grid operations.

The paper studies the use of networked microgrids to manage wildfire ignition risk induced by energized components in the electric network and to promote social equity. We present an optimal microgrid configuration problem for distribution grid operations that limits the risk of wildfire ignitions from electric power lines through preemptive power shutoffs, while prioritizing serving load to socially vulnerable communities. We also present several problem variants to compare various load prioritization formulations and microgrid capabilities. 
Further, we assemble a test case based on a realistic, but not real distribution system that is correlated with actual wildfire risk and social vulnerability data and made available for public use. We demonstrate the efficacy of the proposed method, both on the large-scale realistic test case and through sensitivity studies on a small feeder model. 
We find that the additional flexibility provided by networked microgrids, in combination with our improved modeling of social equity, lead to lower wildfire risk, more load served, and fewer outages in the most socially vulnerable communities. 
Future research may extend this work to consider optimal microgrid design, such as equitable placement of switches and distributed energy resources.

\begin{acks}
This work was authored in part by the National Renewable Energy Laboratory, operated by Alliance for Sustainable Energy, LLC, for the U.S. Department of Energy (DOE) under Contract No. DE-AC36-08GO28308. Funding provided by U.S. Department of Energy Office of Electricity. The views expressed in the article do not necessarily represent the views of the DOE or the U.S. Government. The U.S. Government retains and the publisher, by accepting the article for publication, acknowledges that the U.S. Government retains a nonexclusive, paid-up, irrevocable, worldwide license to publish or reproduce the published form of this work, or allow others to do so, for U.S. Government purposes.
This work is also supported by the National Science Foundation Graduate Research Fellowship Program under Grant No. DGE-1747503. Any opinions, findings, and conclusions or recommendations expressed in this material are those of the authors and do not necessarily reflect the views of the National Science Foundation.
\end{acks}

\bibliographystyle{ACM-Reference-Format}
\bibliography{refs}

%%% -*-BibTeX-*-
%%% Do NOT edit. File created by BibTeX with style
%%% ACM-Reference-Format-Journals [18-Jan-2012].

\begin{thebibliography}{47}

%%% ====================================================================
%%% NOTE TO THE USER: you can override these defaults by providing
%%% customized versions of any of these macros before the \bibliography
%%% command.  Each of them MUST provide its own final punctuation,
%%% except for \shownote{}, \showDOI{}, and \showURL{}.  The latter two
%%% do not use final punctuation, in order to avoid confusing it with
%%% the Web address.
%%%
%%% To suppress output of a particular field, define its macro to expand
%%% to an empty string, or better, \unskip, like this:
%%%
%%% \newcommand{\showDOI}[1]{\unskip}   % LaTeX syntax
%%%
%%% \def \showDOI #1{\unskip}           % plain TeX syntax
%%%
%%% ====================================================================

\ifx \showCODEN    \undefined \def \showCODEN     #1{\unskip}     \fi
\ifx \showDOI      \undefined \def \showDOI       #1{#1}\fi
\ifx \showISBNx    \undefined \def \showISBNx     #1{\unskip}     \fi
\ifx \showISBNxiii \undefined \def \showISBNxiii  #1{\unskip}     \fi
\ifx \showISSN     \undefined \def \showISSN      #1{\unskip}     \fi
\ifx \showLCCN     \undefined \def \showLCCN      #1{\unskip}     \fi
\ifx \shownote     \undefined \def \shownote      #1{#1}          \fi
\ifx \showarticletitle \undefined \def \showarticletitle #1{#1}   \fi
\ifx \showURL      \undefined \def \showURL       {\relax}        \fi
% The following commands are used for tagged output and should be
% invisible to TeX
\providecommand\bibfield[2]{#2}
\providecommand\bibinfo[2]{#2}
\providecommand\natexlab[1]{#1}
\providecommand\showeprint[2][]{arXiv:#2}

\bibitem[TAM(2014)]%
        {TAMU2014}
 \bibinfo{year}{2014}\natexlab{}.
\newblock \bibinfo{title}{How do power lines cause wildfires?}
\newblock
\newblock
\urldef\tempurl%
\url{https://wildfiremitigation.tees.tamus.edu/faqs/how-power-lines-cause-wildfires}
\showURL{%
\tempurl}


\bibitem[sta(2019)]%
        {stanford2019pge}
 \bibinfo{year}{2019}\natexlab{}.
\newblock \bibinfo{title}{PG\&E power outages bring high economic costs for
  California}.
\newblock
\newblock
\urldef\tempurl%
\url{https://woods.stanford.edu/news/pge-power-outages-bring-high-economic-costs-california}
\showURL{%
\tempurl}


\bibitem[pge(2022)]%
        {pgeMitigation}
 \bibinfo{year}{2022}\natexlab{}.
\newblock \bibinfo{title}{2022 Wildfire Mitigation Plan Update}.
\newblock
\newblock
\urldef\tempurl%
\url{https://www.pge.com/en_US/safety/emergency-preparedness/natural-disaster/wildfires/wildfire-mitigation-plan.page?WT.mc_id=Vanity_wildfiremitigationplan}
\showURL{%
\tempurl}


\bibitem[CDC(2022)]%
        {CDCSVI}
 \bibinfo{year}{2022}\natexlab{}.
\newblock \bibinfo{title}{CDC/ATSDR Social Vulnerability Index}.
\newblock
\newblock
\urldef\tempurl%
\url{https://www.atsdr.cdc.gov/placeandhealth/svi/index.html}
\showURL{%
\tempurl}


\bibitem[HHS(2022)]%
        {HHSemPOWER}
 \bibinfo{year}{2022}\natexlab{}.
\newblock \bibinfo{title}{HHS emPOWER Map}.
\newblock
\newblock
\urldef\tempurl%
\url{https://empowerprogram.hhs.gov/about-empowermap.html}
\showURL{%
\tempurl}


\bibitem[Jus(2022)]%
        {Justice40}
 \bibinfo{year}{2022}\natexlab{}.
\newblock \bibinfo{title}{Justice40 initiative}.
\newblock
\newblock
\urldef\tempurl%
\url{https://www.whitehouse.gov/environmentaljustice/justice40/}
\showURL{%
\tempurl}


\bibitem[Cen(2023)]%
        {CensusCRE}
 \bibinfo{year}{2023}\natexlab{}.
\newblock \bibinfo{title}{Community Resilience Estimates}.
\newblock
\newblock
\urldef\tempurl%
\url{https://www.census.gov/programs-surveys/community-resilience-estimates.html}
\showURL{%
\tempurl}


\bibitem[IEE(2023)]%
        {IEEE13orig}
 \bibinfo{year}{2023}\natexlab{}.
\newblock \bibinfo{title}{IEEE PES Test Feeder}.
\newblock
\newblock
\urldef\tempurl%
\url{https://cmte.ieee.org/pes-testfeeders/resources/}
\showURL{%
\tempurl}


\bibitem[Med(2023)]%
        {MedicalBaseline}
 \bibinfo{year}{2023}\natexlab{}.
\newblock \bibinfo{title}{Medical Baseline Program}.
\newblock
\newblock
\urldef\tempurl%
\url{https://www.pge.com/en_US/residential/save-energy-money/help-paying-your-bill/longer-term-assistance/medical-condition-related/medical-baseline-allowance/medical-baseline-allowance.page}
\showURL{%
\tempurl}


\bibitem[WFP(2023)]%
        {WFPI}
 \bibinfo{year}{2023}\natexlab{}.
\newblock \bibinfo{title}{{Wildland Fire Potential Index}}.
\newblock
\newblock
\urldef\tempurl%
\url{https://www.usgs.gov/ecosystems/lcsp/fire-danger-forecast/wind-enhanced-fire-potential-index-wfpi}
\showURL{%
\tempurl}


\bibitem[Amirioun et~al\mbox{.}(2018)]%
        {amirioun2017resilience}
\bibfield{author}{\bibinfo{person}{M.~H. Amirioun}, \bibinfo{person}{F.
  Aminifar}, {and} \bibinfo{person}{H. Lesani}.}
  \bibinfo{year}{2018}\natexlab{}.
\newblock \showarticletitle{Resilience-oriented proactive management of
  microgrids against windstorms}.
\newblock \bibinfo{journal}{\emph{IEEE Trans. Power Syst.}}
  \bibinfo{volume}{33}, \bibinfo{number}{4} (\bibinfo{year}{2018}),
  \bibinfo{pages}{4275--4284}.
\newblock


\bibitem[Anderson and Bell(2012)]%
        {anderson2020}
\bibfield{author}{\bibinfo{person}{G.~B. Anderson} {and} \bibinfo{person}{M~L.
  Bell}.} \bibinfo{year}{2012}\natexlab{}.
\newblock \showarticletitle{{Lights out: impact of the August 2003 power outage
  on mortality in New York, NY.}}
\newblock \bibinfo{journal}{\emph{Epidemiology}} \bibinfo{volume}{23},
  \bibinfo{number}{2} (\bibinfo{year}{2012}), \bibinfo{pages}{189--193}.
\newblock


\bibitem[Arif and Wang(2017)]%
        {arif-NMGs}
\bibfield{author}{\bibinfo{person}{Anmar Arif} {and} \bibinfo{person}{Zhaoyu
  Wang}.} \bibinfo{year}{2017}\natexlab{}.
\newblock \showarticletitle{Networked microgrids for service restoration in
  resilient distribution systems}.
\newblock \bibinfo{journal}{\emph{IET Gener. Transm. Distrib.}}
  \bibinfo{volume}{11}, \bibinfo{number}{14} (\bibinfo{year}{2017}),
  \bibinfo{pages}{3612--3619}.
\newblock


\bibitem[Arnold et~al\mbox{.}(2016)]%
        {arnold2016optimal}
\bibfield{author}{\bibinfo{person}{Daniel~B. Arnold}, \bibinfo{person}{Michael
  Sankur}, \bibinfo{person}{Roel Dobbe}, \bibinfo{person}{Kyle Brady},
  \bibinfo{person}{Duncan~S. Callaway}, {and} \bibinfo{person}{Alexandra
  Von~Meier}.} \bibinfo{year}{2016}\natexlab{}.
\newblock \showarticletitle{Optimal dispatch of reactive power for voltage
  regulation and balancing in unbalanced distribution systems}. In
  \bibinfo{booktitle}{\emph{Proc. IEEE PES General Meeting}}.
  \bibinfo{pages}{1--5}.
\newblock


\bibitem[Astudillo et~al\mbox{.}(2022)]%
        {astudillo2022managing}
\bibfield{author}{\bibinfo{person}{Ayla Astudillo}, \bibinfo{person}{Bai Cui},
  {and} \bibinfo{person}{Ahmed~S. Zamzam}.} \bibinfo{year}{2022}\natexlab{}.
\newblock \showarticletitle{Managing Power Systems-Induced Wildfire Risks Using
  Optimal Scheduled Shutoffs}. In \bibinfo{booktitle}{\emph{Proc. IEEE PES
  General Meeting}}.
\newblock


\bibitem[Brockway et~al\mbox{.}(2021)]%
        {brockway2021inequitable}
\bibfield{author}{\bibinfo{person}{Anna~M Brockway}, \bibinfo{person}{Jennifer
  Conde}, {and} \bibinfo{person}{Duncan Callaway}.}
  \bibinfo{year}{2021}\natexlab{}.
\newblock \showarticletitle{Inequitable access to distributed energy resources
  due to grid infrastructure limits in California}.
\newblock \bibinfo{journal}{\emph{Nature Energy}} \bibinfo{volume}{6},
  \bibinfo{number}{9} (\bibinfo{year}{2021}), \bibinfo{pages}{892--903}.
\newblock


\bibitem[Chen et~al\mbox{.}(2020)]%
        {Chen-NMGs}
\bibfield{author}{\bibinfo{person}{Bo Chen}, \bibinfo{person}{Jianhui Wang},
  \bibinfo{person}{Xiaonan Lu}, \bibinfo{person}{Chen Chen}, {and}
  \bibinfo{person}{Shijia Zhao}.} \bibinfo{year}{2020}\natexlab{}.
\newblock \showarticletitle{Networked microgrids for grid resilience,
  robustness, and efficiency: A review}.
\newblock \bibinfo{journal}{\emph{IEEE Trans. Smart Grid}}
  \bibinfo{volume}{12}, \bibinfo{number}{1} (\bibinfo{year}{2020}),
  \bibinfo{pages}{18--32}.
\newblock


\bibitem[Fobes and Bent(2022)]%
        {PMONM}
\bibfield{author}{\bibinfo{person}{David Fobes} {and} \bibinfo{person}{Russell
  Bent}.} \bibinfo{year}{2022}\natexlab{}.
\newblock \bibinfo{title}{PowerModelsONM.jl}.
\newblock
  \bibinfo{howpublished}{\url{https://github.com/lanl-ansi/PowerModelsONM.jl}}.
\newblock


\bibitem[Fobes et~al\mbox{.}(shed)]%
        {Fobes23}
\bibfield{author}{\bibinfo{person}{David~M Fobes}, \bibinfo{person}{Harsha
  Nagarajan}, {and} \bibinfo{person}{Russell Bent}.} \bibinfo{year}{to be
  published}\natexlab{}.
\newblock \showarticletitle{Optimal Microgrid Networking for Maximal Load
  Delivery in Phase Unbalanced Distribution Grids: A Declarative Modeling
  Approach}.
\newblock \bibinfo{journal}{\emph{IEEE Trans. Smart Grid}} (\bibinfo{year}{to
  be published}).
\newblock


\bibitem[Fuller(2019)]%
        {Fuller2019pge}
\bibfield{author}{\bibinfo{person}{Thomas Fuller}.}
  \bibinfo{year}{2019}\natexlab{}.
\newblock \showarticletitle{PG\&E Outage Darkens Northern California Amid
  Wildfire Threat}.
\newblock \bibinfo{journal}{\emph{The New York Times}} (\bibinfo{date}{Oct}
  \bibinfo{year}{2019}).
\newblock
\showISSN{0362-4331}
\urldef\tempurl%
\url{https://www.nytimes.com/2019/10/09/us/california-power-outage-PGE.html}
\showURL{%
\tempurl}


\bibitem[Gholami et~al\mbox{.}(2019)]%
        {gholami2017proactive}
\bibfield{author}{\bibinfo{person}{Amin Gholami}, \bibinfo{person}{Tohid
  Shekari}, {and} \bibinfo{person}{Santiago Grijalva}.}
  \bibinfo{year}{2019}\natexlab{}.
\newblock \showarticletitle{Proactive management of microgrids for resiliency
  enhancement: An adaptive robust approach}.
\newblock \bibinfo{journal}{\emph{IEEE Trans. Sustain. Energy}}
  \bibinfo{volume}{10}, \bibinfo{number}{1} (\bibinfo{year}{2019}),
  \bibinfo{pages}{470--480}.
\newblock


\bibitem[Ham and Lee(2022)]%
        {Ham2022behavior}
\bibfield{author}{\bibinfo{person}{Youngjib Ham} {and} \bibinfo{person}{Seulbi
  Lee}.} \bibinfo{year}{2022}\natexlab{}.
\newblock \bibinfo{title}{Behavior Analysis of Socially Vulnerable Households
  Responding to Planned Power Shutoffs}.
\newblock
\newblock
\urldef\tempurl%
\url{https://hazards.colorado.edu/mitigation-matters-report/behavior-analysis-of-socially-vulnerable-households-responding-to-planned-power-shutoffs}
\showURL{%
\tempurl}


\bibitem[Inc.(2021)]%
        {Technosylva2021october}
\bibfield{author}{\bibinfo{person}{Technosylva Inc.}}
  \bibinfo{year}{2021}\natexlab{}.
\newblock \bibinfo{title}{October 9-12, 2019 PSPS Event –Wildfire Analysis
  Report}.
\newblock
\newblock
\urldef\tempurl%
\url{https://www.cpuc.ca.gov/consumer-support/psps/technosylva-2019-psps-event-wildfire-risk-analysis-reports}
\showURL{%
\tempurl}


\bibitem[{Jazebi} et~al\mbox{.}(2020)]%
        {jazebi2020}
\bibfield{author}{\bibinfo{person}{S. {Jazebi}}, \bibinfo{person}{F. {de
  León}}, {and} \bibinfo{person}{A. {Nelson}}.}
  \bibinfo{year}{2020}\natexlab{}.
\newblock \showarticletitle{{Review of Wildfire Management Techniques—Part I:
  Causes, Prevention, Detection, Suppression, and Data Analytics}}.
\newblock \bibinfo{journal}{\emph{IEEE Trans. Power Delivery}}
  \bibinfo{volume}{35}, \bibinfo{number}{1} (\bibinfo{year}{2020}),
  \bibinfo{pages}{430--439}.
\newblock


\bibitem[Kody et~al\mbox{.}(2022a)]%
        {kody2022optimizing}
\bibfield{author}{\bibinfo{person}{Alyssa Kody}, \bibinfo{person}{Ryan
  Piansky}, {and} \bibinfo{person}{Daniel~K. Molzahn}.}
  \bibinfo{year}{2022}\natexlab{a}.
\newblock \bibinfo{title}{Optimizing Transmission Infrastructure Investments to
  Support Line De-energization for Mitigating Wildfire Ignition Risk}.
\newblock
\newblock
\urldef\tempurl%
\url{https://doi.org/10.48550/ARXIV.2203.10176}
\showDOI{\tempurl}


\bibitem[Kody et~al\mbox{.}(2022b)]%
        {kody2022sharing}
\bibfield{author}{\bibinfo{person}{Alyssa Kody}, \bibinfo{person}{Amanda West},
  {and} \bibinfo{person}{Daniel~K. Molzahn}.} \bibinfo{year}{2022}\natexlab{b}.
\newblock \showarticletitle{Sharing the Load: Considering Fairness in
  De-energization Scheduling to Mitigate Wildfire Ignition Risk using Rolling
  Optimization}. In \bibinfo{booktitle}{\emph{Proc. IEEE Conf. Decision
  Control}}. \bibinfo{pages}{5705--5712}.
\newblock


\bibitem[Kroposki et~al\mbox{.}(2020)]%
        {kroposki2020autonomous}
\bibfield{author}{\bibinfo{person}{Benjamin Kroposki}, \bibinfo{person}{Andrey
  Bernstein}, \bibinfo{person}{Jennifer King}, \bibinfo{person}{Deepthi
  Vaidhynathan}, \bibinfo{person}{Xinyang Zhou}, \bibinfo{person}{Chin-Yao
  Chang}, {and} \bibinfo{person}{Emiliano Dall’Anese}.}
  \bibinfo{year}{2020}\natexlab{}.
\newblock \showarticletitle{Autonomous energy grids: Controlling the future
  grid with large amounts of distributed energy resources}.
\newblock \bibinfo{journal}{\emph{IEEE Power Energy Mag.}}
  \bibinfo{volume}{18}, \bibinfo{number}{6} (\bibinfo{year}{2020}),
  \bibinfo{pages}{37--46}.
\newblock


\bibitem[Maeda(2019)]%
        {Maeda_2019}
\bibfield{author}{\bibinfo{person}{John Maeda}.}
  \bibinfo{year}{2019}\natexlab{}.
\newblock \showarticletitle{Design in Tech Report 2019}.
\newblock  (\bibinfo{year}{2019}).
\newblock
\urldef\tempurl%
\url{https://designintech.report/wp-content/uploads/2019/03/dit2019_v00.pdf}
\showURL{%
\tempurl}


\bibitem[Mateo~Domingo et~al\mbox{.}(2011)]%
        {domingo2011reference}
\bibfield{author}{\bibinfo{person}{Carlos Mateo~Domingo},
  \bibinfo{person}{Tomas Gomez San~Roman}, \bibinfo{person}{Alvaro
  Sanchez-Miralles}, \bibinfo{person}{Jesus~Pascual Peco~Gonzalez}, {and}
  \bibinfo{person}{Antonio Candela~Martinez}.} \bibinfo{year}{2011}\natexlab{}.
\newblock \showarticletitle{A Reference Network Model for Large-Scale
  Distribution Planning With Automatic Street Map Generation}.
\newblock \bibinfo{journal}{\emph{IEEE Trans. Power Syst.}}
  \bibinfo{volume}{26}, \bibinfo{number}{1} (\bibinfo{date}{Feb}
  \bibinfo{year}{2011}), \bibinfo{pages}{190–197}.
\newblock
\showISSN{0885-8950, 1558-0679}
\urldef\tempurl%
\url{https://doi.org/10.1109/TPWRS.2010.2052077}
\showDOI{\tempurl}


\bibitem[Mitsova et~al\mbox{.}(2018)]%
        {Mitsova_Esnard_Sapat_Lai_2018}
\bibfield{author}{\bibinfo{person}{Diana Mitsova},
  \bibinfo{person}{Ann-Margaret Esnard}, \bibinfo{person}{Alka Sapat}, {and}
  \bibinfo{person}{Betty~S. Lai}.} \bibinfo{year}{2018}\natexlab{}.
\newblock \showarticletitle{Socioeconomic vulnerability and electric power
  restoration timelines in Florida: the case of Hurricane Irma}.
\newblock \bibinfo{journal}{\emph{Natural Hazards}} \bibinfo{volume}{94},
  \bibinfo{number}{2} (\bibinfo{date}{Nov} \bibinfo{year}{2018}),
  \bibinfo{pages}{689–709}.
\newblock
\showISSN{1573-0840}
\urldef\tempurl%
\url{https://doi.org/10.1007/s11069-018-3413-x}
\showDOI{\tempurl}


\bibitem[Mohler(2019)]%
        {mohler2019cal}
\bibfield{author}{\bibinfo{person}{Michael Mohler}.}
  \bibinfo{year}{2019}\natexlab{}.
\newblock \bibinfo{title}{{CAL FIRE Investigators Determine Cause of the Camp
  Fire}}.
\newblock
\newblock
\urldef\tempurl%
\url{https://www.fire.ca.gov/media/5121/campfire_cause.pdf}
\showURL{%
\tempurl}


\bibitem[Moon(2022)]%
        {Moon2022california}
\bibfield{author}{\bibinfo{person}{Sarah Moon}.}
  \bibinfo{year}{2022}\natexlab{}.
\newblock \bibinfo{title}{California’s second-largest wildfire was sparked
  when power lines came in contact with a tree, Cal Fire says}.
\newblock
\newblock
\urldef\tempurl%
\url{https://www.cnn.com/2022/01/05/us/dixie-fire-power-lines-cause-pge/index.html}
\showURL{%
\tempurl}


\bibitem[on~Environmental~Quality(2022)]%
        {CEJST}
\bibfield{author}{\bibinfo{person}{Council on Environmental~Quality}.}
  \bibinfo{year}{2022}\natexlab{}.
\newblock \bibinfo{title}{Climate \& Economic Justice Screening Tool}.
\newblock
\newblock
\urldef\tempurl%
\url{https://screeningtool.geoplatform.gov/en/#3/33.47/-97.5}
\showURL{%
\tempurl}


\bibitem[Palmintier(2023)]%
        {SMART-DS}
\bibfield{author}{\bibinfo{person}{Bryan Palmintier}.}
  \bibinfo{year}{2023}\natexlab{}.
\newblock \bibinfo{booktitle}{\emph{SMART-DS: Synthetic Models for Advanced,
  Realistic Testing: Distribution Systems and Scenarios}}.
\newblock
\urldef\tempurl%
\url{https://www.nrel.gov/grid/smart-ds.html}
\showURL{%
\tempurl}


\bibitem[Rhodes et~al\mbox{.}(2021)]%
        {rhodes2021balancing}
\bibfield{author}{\bibinfo{person}{Noah Rhodes}, \bibinfo{person}{Lewis
  Ntaimo}, {and} \bibinfo{person}{Line~A. Roald}.}
  \bibinfo{year}{2021}\natexlab{}.
\newblock \showarticletitle{Balancing Wildfire Risk and Power Outages Through
  Optimized Power Shut-Offs}.
\newblock \bibinfo{journal}{\emph{IEEE Trans. Power Syst.}}
  \bibinfo{volume}{36}, \bibinfo{number}{4} (\bibinfo{year}{2021}),
  \bibinfo{pages}{3118--3128}.
\newblock


\bibitem[Rhodes and Roald(2021)]%
        {rhodes2021role}
\bibfield{author}{\bibinfo{person}{Noah Rhodes} {and} \bibinfo{person}{Line
  Roald}.} \bibinfo{year}{2021}\natexlab{}.
\newblock \showarticletitle{The Role of Distributed Energy Resources in
  Distribution System Restoration}.
\newblock  (\bibinfo{year}{2021}).
\newblock


\bibitem[Rhodes and Roald(2022)]%
        {rhodes2022cooptimization}
\bibfield{author}{\bibinfo{person}{Noah Rhodes} {and} \bibinfo{person}{Line
  Roald}.} \bibinfo{year}{2022}\natexlab{}.
\newblock \bibinfo{title}{Co-optimization of power line shutoff and restoration
  under high wildfire ignition risk}.
\newblock
\newblock
\urldef\tempurl%
\url{https://doi.org/10.48550/ARXIV.2204.02507}
\showDOI{\tempurl}


\bibitem[Russell et~al\mbox{.}(2012)]%
        {russell2012distribution}
\bibfield{author}{\bibinfo{person}{B.~Don Russell}, \bibinfo{person}{Carl~L.
  Benner}, {and} \bibinfo{person}{Jeffrey~A. Wischkaemper}.}
  \bibinfo{year}{2012}\natexlab{}.
\newblock \showarticletitle{Distribution feeder caused wildfires: Mechanisms
  and prevention}. In \bibinfo{booktitle}{\emph{Proc. 65th Ann. Conf.
  Protective Relay Eng.}} \bibinfo{pages}{43--51}.
\newblock


\bibitem[Sotolongo et~al\mbox{.}(2020)]%
        {Sotolongo2020california}
\bibfield{author}{\bibinfo{person}{Marisa Sotolongo}, \bibinfo{person}{Cecelia
  Bolon}, {and} \bibinfo{person}{Shalanda~H Baker}.}
  \bibinfo{year}{2020}\natexlab{}.
\newblock \showarticletitle{California Power Shutoffs: Deficiencies in Data and
  Reporting}.
\newblock  (\bibinfo{date}{October} \bibinfo{year}{2020}).
\newblock
\urldef\tempurl%
\url{https://bit.ly/3wfPibX}
\showURL{%
\tempurl}


\bibitem[Sotolongo et~al\mbox{.}(2021)]%
        {Sotolongo_Kuhl_Baker_2021}
\bibfield{author}{\bibinfo{person}{Marisa Sotolongo}, \bibinfo{person}{Laura
  Kuhl}, {and} \bibinfo{person}{Shalanda~H. Baker}.}
  \bibinfo{year}{2021}\natexlab{}.
\newblock \showarticletitle{Using environmental justice to inform disaster
  recovery: Vulnerability and electricity restoration in Puerto Rico}.
\newblock \bibinfo{journal}{\emph{Environmental Science \& Policy}}
  \bibinfo{volume}{122} (\bibinfo{year}{2021}), \bibinfo{pages}{59–71}.
\newblock
\showISSN{1462-9011}
\urldef\tempurl%
\url{https://doi.org/10.1016/j.envsci.2021.04.004}
\showDOI{\tempurl}


\bibitem[Syphard and Keeley(2015)]%
        {syphard2015location}
\bibfield{author}{\bibinfo{person}{Alexandra~D Syphard} {and}
  \bibinfo{person}{Jon~E Keeley}.} \bibinfo{year}{2015}\natexlab{}.
\newblock \showarticletitle{Location, timing and extent of wildfire vary by
  cause of ignition}.
\newblock \bibinfo{journal}{\emph{Int. J. Wildland Fire}} \bibinfo{volume}{24},
  \bibinfo{number}{1} (\bibinfo{year}{2015}), \bibinfo{pages}{37--47}.
\newblock


\bibitem[Taylor and Roald(2022)]%
        {taylor2021framework}
\bibfield{author}{\bibinfo{person}{Sofia Taylor} {and} \bibinfo{person}{Line~A
  Roald}.} \bibinfo{year}{2022}\natexlab{}.
\newblock \showarticletitle{A Framework for Risk Assessment and Optimal Line
  Upgrade Selection to Mitigate Wildfire Risk}.
\newblock \bibinfo{journal}{\emph{Electr. Power Syst. Res}}
  (\bibinfo{year}{2022}).
\newblock


\bibitem[Taylor et~al\mbox{.}(2023)]%
        {OptMGConfigRepo}
\bibfield{author}{\bibinfo{person}{Sofia Taylor}, \bibinfo{person}{Gabriela
  Setyawan}, \bibinfo{person}{Bai Cui}, \bibinfo{person}{Ahmed Zamzam}, {and}
  \bibinfo{person}{Line~A. Roald}.} \bibinfo{year}{2023}\natexlab{}.
\newblock \bibinfo{booktitle}{\emph{{Optimal-Microgrid-Configuration}}}.
\newblock
\urldef\tempurl%
\url{https://github.com/WISPO-POP/Optimal-Microgrid-Configuration}
\showURL{%
\tempurl}


\bibitem[Teague et~al\mbox{.}(2010)]%
        {victoria2009final}
\bibfield{author}{\bibinfo{person}{Bernard~George Teague},
  \bibinfo{person}{Ronald~N McLeod}, {and} \bibinfo{person}{Susan~Mary
  Pascoe}.} \bibinfo{year}{2010}\natexlab{}.
\newblock \bibinfo{booktitle}{\emph{Final Report: 2009 Victorian Bushfires
  Royal Commission}}.
\newblock \bibinfo{publisher}{Victorian Bushfires Royal Commission, Australia}.
\newblock
\newblock
\shownote{\url{http://royalcommission.vic.gov.au/Commission-Reports/Final-Report.html},
  last accessed April 2020}.


\bibitem[Tormos-Aponte et~al\mbox{.}(2021)]%
        {Tormos2021}
\bibfield{author}{\bibinfo{person}{Fernando Tormos-Aponte},
  \bibinfo{person}{Gustavo García-López}, {and}
  \bibinfo{person}{Mary~Angelica Painter}.} \bibinfo{year}{2021}\natexlab{}.
\newblock \showarticletitle{Energy inequality and clientelism in the wake of
  disasters: From colorblind to affirmative power restoration}.
\newblock \bibinfo{journal}{\emph{Energy Policy}}  \bibinfo{volume}{158}
  (\bibinfo{year}{2021}), \bibinfo{pages}{112550}.
\newblock
\showISSN{0301-4215}
\urldef\tempurl%
\url{https://doi.org/10.1016/j.enpol.2021.112550}
\showDOI{\tempurl}


\bibitem[Wang et~al\mbox{.}(2016)]%
        {Wang-NMGs}
\bibfield{author}{\bibinfo{person}{Zhaoyu Wang}, \bibinfo{person}{Bokan Chen},
  \bibinfo{person}{Jianhui Wang}, {and} \bibinfo{person}{Chen Chen}.}
  \bibinfo{year}{2016}\natexlab{}.
\newblock \showarticletitle{Networked Microgrids for Self-Healing Power
  Systems}.
\newblock \bibinfo{journal}{\emph{IEEE Trans. Smart Grid}} \bibinfo{volume}{7},
  \bibinfo{number}{1} (\bibinfo{year}{2016}), \bibinfo{pages}{310--319}.
\newblock


\bibitem[Yang et~al\mbox{.}(2022)]%
        {yang2022resilient}
\bibfield{author}{\bibinfo{person}{Weijia Yang}, \bibinfo{person}{Sarah~N.
  Sparrow}, \bibinfo{person}{Masaō Ashtine}, \bibinfo{person}{David~C.H.
  Wallom}, {and} \bibinfo{person}{Thomas Morstyn}.}
  \bibinfo{year}{2022}\natexlab{}.
\newblock \showarticletitle{Resilient by design: Preventing wildfires and
  blackouts with microgrids}.
\newblock \bibinfo{journal}{\emph{Applied Energy}}  \bibinfo{volume}{313}
  (\bibinfo{year}{2022}), \bibinfo{pages}{118793}.
\newblock
\showISSN{0306-2619}
\urldef\tempurl%
\url{https://doi.org/10.1016/j.apenergy.2022.118793}
\showDOI{\tempurl}


\end{thebibliography}

\end{document}